\documentclass[journal]{IEEEtran}
\usepackage[utf8]{inputenc}

\usepackage{caption} 
\usepackage{color}
\usepackage{amsmath}
\usepackage{graphicx}
\definecolor{purple}{RGB}{204, 0, 255}	
\definecolor{maroon}{RGB}{176, 48, 96}

\definecolor{green}{RGB}{79, 200, 66}

\definecolor{blue}{RGB}{0,0,255}

\title{Opinion Diffusion Software with Strategic Opinion Revelation and Unfriending}
\author{\IEEEauthorblockN{Patrick Shepherd, Mia Weaver, and Judy Goldsmith} 
\IEEEauthorblockA{Department of Computer Science, University of Kentucky, Lexington, KY, 40506-0633, USA\\
Email:  patrick.shepherd@uky.edu and mia.weaver@uky.edu and goldsmit@cs.uky.edu
}
} 

\begin{document}

\maketitle

\begin{abstract}
We present a novel software suite for social network modeling and opinion diffusion processes.  Much research on social network science has assumed networks with static topologies.  More recently, attention has been turned to networks that evolve.  Although software for modeling both the topological evolution of networks and diffusion processes are constantly improving, very little attention has been paid to agent modeling.  Our software is designed to be robust, modular, and extensible, providing the ability to model dynamic social network topologies and multidimensional diffusion processes, different styles of agent including non-homophilic paradigms, as well as a testing environment for multi-agent reinforcement learning (MARL) experiments with diverse sets of agent types.  We also illustrate the value of diverse agent modeling, and environments that allow for strategic unfriending.  Our work shows that polarization and consensus dynamics, as well as topological clustering effects, may rely more than previously known on individuals' goals for the composition of their neighborhood's opinions.
\end{abstract}

\section{Introduction}
The study of 
social networks
has increased dramatically in recent years as their ubiquity impacts virtually every aspect of our lives.  One of the main topics of discussion is that of opinion manipulation over social networks.  We have built a framework that includes most 
mechanics common in the social network simulation and opinion diffusion literature.  The goal of this framework is to provide a robust and flexible platform for rich opinion- and topology-centered studies, as well as a test bed for reinforcement learning agents.  A key feature of our platform is the inclusion of node \emph{archetypes} --- categories that determine what an individual in a network values.  These archetypes determine how each node behaves within the network.

Current events have brought significant attention to the ability of online social interaction to effect outcomes in the real world.  It is postulated that bad actors were able to sway the direction of both the Brexit vote in the 
UK and the 
US presidential election of 2016 (e.g., \cite{sanger2019russia}).  Even outside the realm of major historical events, discussion abounds of the effects --- both good and bad --- of selective exposure to information in online social networks, running the gamut from self-realization to radicalization.  Whatever the true extent of these effects is, it has become hard to deny that they exist.

Research on diffusion and other processes over social networks has taken many directions.  Classical models viewed social networks as nodes connected by undirected edges, with a single binary state for each node 
e.g.,
\emph{active} and \emph{inactive} \cite{french1956formal,granovetter1978threshold,kempe2003maximizing}.  These models simulate some condition spreading through the network in discrete time steps, given a single rule to determine each node's state at each time step based in some way on the states of one or more of the node's neighbors.  Research building on these models added features including: social influence, making some nodes more effective at propagating their state than others \cite{baumann2020modeling,DeGroot,friedkin2016network,weisbuch2002meet}; masks, enabling nodes to either broadcast their state to the network or keep it hidden \cite{grandi2016strategic}; multidimensional diffusion spaces, which allow more than one phenomenon to spread through the network simultaneously \cite{chen2018modeling,DeGroot,friedkin2016network}; and in the multidimensional case, correlations between issues so that a node is more or less likely to flip their state in one dimension depending on its state in another \cite{dellaposta2015liberals,friedkin2016network}.

Most research in opinion diffusion rests on the assumption that all agents in a network prefer connections to others who hold the same opinion(s); there are exceptions to this rule, and some forms of opinion antagonism and their effects on the topology and opinion space of social networks have occasionally been investigated \cite{kurmyshev2011dynamics,motsch2014heterophilious,sirbu2013opinion}.

\section{Related Work}

The present paper has two main thrusts: the introduction of our modeling software 
and a series of opinion diffusion experiments using it.  Here we briefly discuss the literature relevant to each.

\noindent \textbf{Social Network Modeling Software}

Stadtfeld's R package \emph{NetSim} \cite{stadtfeld2015netsim} is a simulation software largely focused on topological social network simulations, but also provides facilities for network simulations involving arbitrary attribute revisions over time.\footnote{This software allows for arbitrary attributes to be associated with individual nodes (e.g., gender, height, opinion, etc.)  Users may also program their own network change mechanism, by which agents update their attribute values based on a user-defined rule.  However, there is no built-in functionality to, e.g., embed nodes in an opinion space.}  The software includes: 
different time models (continuous or discrete time, round-based time scales, etc.); behavior models, which can be applied to all agents in the network or just subsets of them; and models to change the state of the network with each update.  This software is extremely flexible, but is also very general, leaving a great deal of the modeling responsibility in the hands of the user, as does the software presented here.  However, our package features a prebuilt architecture for opinion diffusion processes in arbitrary dimensions, agent types that conform to most current simulation needs, and aggregation procedures.  We also provide functionality for the adaptation of our environment to reinforcement learning applications.

Another network diffusion simulation software is NDLib \cite{rossetti2017ndlib}.  The software is designed to provide facilities specifically for diffusive phenomena such as epidemics and opinion spread, and could be applied to virtually any diffusive process, and there are a number of common diffusion models already supported by the software.  However, Rosetti \emph{et al.} don't describe opportunities for agent modeling in NDLib.

Ryczko \emph{et al.} \cite{ryczko2017hashkat} developed \emph{Hashkat}, a large-scale network evolution simulator.  The software is mainly meant to simulate the growth over time of a social network by allowing for the definition of the rates at which certain events happen.  The software models information propagation, but does not include facilities for updating agents' positions within any attribute space.  However, it does include the ability to create different agent types with respect to features such as the ratio of in- to out-degree, overall connectivity, and overall proportional representation within the network.

Other software packages abound that perform some network simulation capabilities, such as the R package \emph{iGraph} \cite{csardi2006igraph}, Gephi\footnote{https://gephi.org/} and GraphViz\footnote{https://www.graphviz.org/} for visualization purposes, and the Python package Nepidemix\footnote{http://nepidemix.irmacs.sfu.ca/}, built on top of NetworkX, for simulating the spread of an epidemic.  Chuan \emph{et al.}~\cite{chuan2018design} have also developed a platform for conducting 
phenomenon diffusion experiments on 
large networks.

\noindent \textbf{Agent Based Modeling of Opinion Diffusion}

Pilditch \emph{et al.}~\cite{pilditch2017opinion} developed an agent-based opinion cascade model in which opinions diffuse through a network based on individual agent decisions.  The authors develop synthetic networks in a unidimensional opinion space, with agents moving through opinion space based on their observations of their friends' known opinions.  All agents begin the simulation with a neutral attitude toward both opinions, and all use the same model to determine their opinion updates.  Each step, agents have the option to declare their opinion.  Then agents observe the opinion of the first of their neighbors to declare, and use a simple Q-learning model to determine whether or not they will update their own opinions.  Their experiments on a uniform agent type showed that the network stayed nearly evenly split in terms of opinions, but that clustering with like-minded agents was a dependable outcome.

Duggins \cite{duggins2014psychologically} provides a similar model.  In it, agents have attributes for tolerance of dissimilar opinion, susceptibility to social influence, and desire to conform to social norms, where ``conformity" refers to how far between an agent's true opinion and the socially normative opinion the agent wishes to appear to others.  The model allows users a single opinion on a 0--100 spectrum, and has a static topology.  Simulations involve individuals in neighborhoods expressing an opinion somewhere between what they truly believe and the socially normative opinion. 
Ye \emph{et al.}~\cite{ye2019influence} also investigate a model featuring separate private and public opinions for agents.

Madsen \emph{et al.}~\cite{madsen2018large} use a similar model, but with an 
update scheme roughly the same as the bounded confidence model \cite{deffuant2000mixing}.  Agents have 
a real-valued opinion about the state of the world.
Each time step agents seek out others whose opinion is close enough to their own, then update their opinion based on the aggregation of observed opinions.  Agents' confidence in their opinions also changes over time.  Toscani \emph{et al.} \cite{toscani2018opinion} use a ``kinetic" model of opinion formation, or one in which both agents' prior opinion and connectivity determine the outcome of the opinion update process.  


Chen \emph{et al.} \cite{chen2018modeling} model agent-based opinion dynamics with an additional personality parameters, building even further on the notion that individuals with similar personalities will tend to form stronger ties with each other.  They use multidimensional opinion space, 
with
an opinion velocity factor to maintain 
naturalistic fluidity in the model.  New edges are created between pairs of nodes with probability proportional to the Euclidean distance between their opinion vectors.  Li \emph{et al.} \cite{li2017simulation} use the Stubborn Individuals and Orators (SO) model \cite{di2007opinion}, which uses two additional parameters to model how resistant individuals are to opinion change, and how influential individuals are, both of which our model accommodates.  
Banisch \emph{et al.} \cite{banisch2019opinion} explore the dynamics of opinion formation when social feedback is used as a reinforcement learning signal.  

We
direct the reader to \cite{anderson2019recent} for a more complete overview of opinion diffusion techniques and models.  Although some research considers agent heterogeneity in the domains of opinion, resistance to opinion change, and interpersonal influence, none to our knowledge model agents with goals that are not necessarily homophilic.  The contributions of our work are a novel software suite for diffusion modeling with a focus on differing agent types, and a first investigation of network outcomes when agents differ in both their preferences over distance from their neighbors in opinion space and the way in which observed opinions change their own.

\section{The Software:} Our software is designed to be modular, general, and easily extensible, while providing many
features from the phenomenon diffusion literature.  Our platform is written in Python, relying mainly on NetworkX \cite{networkX} for its network generation facilities, and its extensive collection of standard graph-theoretic metrics and community detection algorithms.  We also employ matplotlib\footnote{https://matplotlib.org/} for some visualizations of metrics and networks, and GraphViz for more detailed network visualizations.

\subsection{Network Generation and Metric Collection:} Like NetSim, almost every aspect of our software is easily customizable.  First, a number of statistics collection functions have already been included and are ready to use.  Due to the compatibility with NetworkX, that library's entire collection of metric calculation functions can be incorporated by adding a small wrapper function to the SocialNetworkGraph class.  Further, network generation algorithms are just as easy to include.  Our software comes prepackaged with the ability to generate graphs via one of three algorithms: random \cite{erdds1959random,gilbert1959random}, small world \cite{watts1998collective}, and scale-free \cite{barabasi1999emergence,dorogovtsev2000structure}.  Once a network is built, a mechanism to define the process of creating or destroying edges over time can be easily customized.  By default, the software probabilistically adds links to create new triadic closures in the network at each time step, and the removal of edges is the purview of agents' behavioral models, described below.

\subsection{Agent and Edge Properties:} Each node in the network has a list of characteristic default properties that have some effect over its position and status in the network over time.  New attributes can easily be added and taken into account during simulations.

    $\bullet$\  $\vec{b}_i^t = \{ -1, 1 \}^K$ : $i$'s \emph{private opinion} on each of $K$ topics at time $t$,\footnote{For this and following time-dependent attributes, we will exclude the superscript $t$ where unnecessary.} and we refer to $i$'s opinion on topic $k$ as $b_{ik}^t$,
    
    $\bullet$\  $\mathit{r}^*_i$ : the reward function for $i$'s \emph{archetype} (*), or utility derived from the known opinions in $N(i)$,
    
    $\bullet$\  $\mathit{upd}_i$ : a rule dictating the criterion by which $i$ will change an opinion in $\vec{b}_i$,
    
    $\bullet$\  $\Pr( \mathit{upd}_i )$ : the probability an agent will change its opinion in the face of disagreement (or agreement) depending on its archetype,
    
    $\bullet$\  $\Pr(\mathit{unf})$ : the probability that a node will sever its connection to another node if the connection is not valuable enough,
    
    $\bullet$\  $res(i)$ : $i$'s degree of resistance to opinion influence,
    
    $\bullet$\  $A(i)$ : the set of actions available to $i$ at each time step,

The properties associated with edges  govern interactions between agents.

    $\bullet$\  $w_{ij} \in [0,1]$ : the \emph{weight} associated with the edge from $i$ to $j$, to be understood as $i$'s influence over $j$.  The column vector $\vec{w}_{*i} = \{ w_{ji} : j \in V \}$ represents all influence exerted over $i$.  All $\vec{w}_{i}$ are columns in a matrix $\mathbf{W}$.  We will refer to a normalized version of this matrix as described below for all relevant calculations.
    
    $\bullet$\  $\vec{m}_{ij}^t = \{ -1, 0, 1 \}^K$ : a \emph{masking vector} describing which of $i$'s opinions are revealed to $j$ at time $t$.  Entry $m_{ijk}^t = 0$ if $i$ hasn't revealed its opinion on topic $k$ to $j$ by time $t$, and $b_{ik}^t$ otherwise.  All $\vec{m}_{ij}^t$ are rows of matrix $\mathbf{m}_i^t$ 
    containing
    information about which opinions $i$ has revealed to each other 
    neighbor.
    For any $j$ s.t. $(i,j) \notin E$ at time $t$, $\vec{m}_{ij}^t = 0^K$.  Therefore, if $i$ unfriends $j$ and the network later re-friends them at time $t$, then $m_{ijk}^t = 0=m_{jik}^t$ for all $K$ issues.

\subsection{Programmable Features:} A number of mechanisms can be manipulated or created anew for simulations.  The foremost of these is agent archetypes.  Each agent in the network can be outfitted with a custom archetype, or series of rules and characteristics that determine the agent's behavior in different situations.  An agent's archetype comprises three aspects: their \emph{reward}, or satisfaction with their status in the network at a given point in time; their \emph{update rule}, determining in what cases the agent will change its opinion on a topic; and their \emph{policy}, the set of conditions determining what action the agent will take.

In our platform, the reward an agent gets at a specific time step is by default, although not necessarily, related to the distance in opinion space between itself and its neighbors.  In an opinion diffusion process with completely visible opinions, this reduces to calculating a similarity metric between $\vec{b}_i$ and $\vec{b}_{j}$ for $(i, j) \in E$ (see below).

The update rule for each agent is a function taking a neighborhood's average opinion (the calculation of which is also built in, but can be manipulated) as input and producing a single agent's new opinion vector.  In 
most
models, this rule involves moving agents closer to each other in opinion space rather than farther away.  
An update rule is of the form:
\textbf{if} \emph{neighborhood average opinion is strong enough} \textbf{and} \emph{opposite of what I want}, \textbf{then} \emph{flip opinion with probability} $\Pr(upd_i)$."  The threshold for ``strong enough" and the definition of ``what I want" are customizable.  Finally, an agent's policy determines its actions given its state at a moment in time.  Our software comes prepackaged with simple opinion-revelation and edge-deletion policies that can be manipulated as needed.  Additionally, we have provided some facilities to link agent behavior to a learning model such as a neural network, and have the policy provided by that.

The software provides agents with a menu of actions at each time step from which they can choose.  Our software was 
conceived as a reinforcement learning platform for strategic opinion revelation and unfriending, so the actions available are: a) $\mathit{reveal}(i,j,k)$ --- $i$ reveals opinion $k$ to $j$; b) $\mathit{unfriend}(i,j)$ --- $i$ unfriends $j$; and c) NOP.  This set of actions can be expanded or replaced as needed for any given simulation.

It will often be the case that there is no need in a particular setting to model different types of agents with different goals, though, and so creating each of these features for a single global agent type provides a significant amount of flexibility in the range of experiments that can be performed.


\subsection{Simulation Flow} A simulation in our framework 
has
four steps: 1) agents observe their neighbors' states and determine actions to take with each, 2) each agent executes its chosen actions, 3) each agent updates its opinion according to its update rule, and 4) new friend connections are introduced by the network itself.

There are 
many
approaches used to model diffusive processes in a network: only allowing a single agent to perform a single action, allowing all agent to perform exactly one action, and most 
variants
in between.  Our software defaults to allowing each agent to choose an action \emph{for each of its neighbors}, though this is an easy mechanism to relax if desired.  
Agent $i$ 
chooses an action w.r.t.
$j$ based on what it knows about $j$'s opinions, by observing $\vec{m}_{ji}$, and calculating the distance between that and its own opinion vector.  The default 
distance metric is introduced in the next section.  
The 
the policy $i$ uses to guide its actions can be defined as needed. 
A simple policy 
could
take the 
form: ``If I dislike my neighbor's opinion enough, I will probably unfriend them.  Otherwise, I may reveal an opinion to them to 
bolster our connection, or 
do nothing 
for now."
The actions a node takes ultimately 
affects
both the topology and the overall opinion space of the network.  The friending/unfriending mechanism allows 
agents to directly control 
their local network topology, to improve their own satisfaction with their neighborhood.
Further, if $i$ reveals an opinion to a disagreeing neighbor $j$, it may cause $j$ to flip its opinion, potentially resulting in higher reward for $i$.  The dynamics of strategic revelation could have significant effects on overall network outcomes, which is a main subject of our ongoing work.

Once actions have been chosen and executed, each agent updates its opinion based on its new environment as it may have some new neighbors and be missing some former ones, and the opinions of remaining neighbors may have changed.  We denote the complete set of private opinions, or \emph{opinion profile}, at time $t$ as $\mathbf{B}^t = \{ \vec{b}_i^t : i \in V \}$.  At each time step, agents are able to see only the opinions that their neighbors have revealed to them and use those to update their beliefs according to their archetype.  We will refer to what $i$ knows about its neighbors' opinions at time $t$ as $i$'s \emph{view:} $\widehat{\mathbf{b}}_i^t = \bigcup_{j \in V} \vec{m}_{ji}^t$.  At the beginning of each time step, an agent must determine the \emph{aggregate opinion} over its neighborhood by calculating the average known opinion over its friends, weighted by their proportional contribution to the agent's incoming influence.  Let $\overline{\mathbf{W}}$ be a matrix in which $\overline{\mathbf{W}} = \overline{w}_{ji} = w_{ji} / \sum_{j \in N(i)} w_{ji}$.  Then we can define the \emph{aggregate opinion profile} within $i$'s neighborhood as

\begin{equation}\label{aggregate}
    \widetilde{\mathbf{b}}^t_i = \overline{\mathbf{W}}_{* i}^\top \widehat{\mathbf{b}}_i
\end{equation}
and the average opinion on topic $k$ within $i$'s neighborhood is $\widetilde{b}_{ik}$.  Note that, when all weights are 
equal, this reduces the calculation of the neighborhood average opinion on topic $k$ to $\widetilde{b}_{ik} = |N(i)|^{-1} \sum_{j \in N(i)} m_{jik}$.  With this, we can determine whether $i$'s neighborhood mostly agrees with 
$i$
or not by testing $\widetilde{b}_{ik} * b_{ik} < 0$ or not.  This information can then be used to determine how $i$'s opinions move through opinion space.  Almost all relevant research encompasses agents that only move closer to their neighbors in opinion space, but some attention has been paid to other paradigms \cite{kurmyshev2011dynamics,motsch2014heterophilious,sirbu2013opinion}.

\subsection{Default Settings}Our software is designed to be general and thus capable of supporting a wide range of subject matter in experiments, but is equipped specifically for simulations that monitor the topology and opinion space of a social network (and changes thereto).  To that end, it includes several default settings that emulate widely-used configurations.  Here we describe some components that are ready to be used off-the-shelf.

In a simulation, agents must first choose actions to take at each time step with 
some or
all of their neighbors.  When 
choosing an
action to take with a particular neighbor, agents assess the 
reward
they get from the connection, based on their archetype.  
For instance, if $i$ decides its reward is being harmed too much by its connection to $j$, it may choose to sever the edge connecting them.  Otherwise, $i$ 
might
reveal a new opinion to $j$ if they are known to agree on other issues already.

Agent reward is based on \emph{distance} in opinion space, and our default calculation of distance between agents is based on the agreement between known opinions only, relative to one of the agents.  In other words, to find the distance $i$ perceives itself to be from $j$, we first take the set of issues on which $j$ has revealed its opinion to $i$, $ \vec{v}_{ij} = \{ b_{jk} : m_{jik} \neq 0 \}$.   Let $n_{ij} = |\vec{v}_{ij}|$ be the number of issues; then distance $d(i,j)$ is the number of issues in $ \vec{v}_{ij} $ on which $\vec{b}_i$ and $\vec{b}_j$ disagree, divided by $n_{ij}$.


There are two notions of reward in our software: agent-to-agent reward, and neighborhood-to-agent reward.  Let the agent-to-agent reward $i$ gets from $j$ be defined by a function $r^*_i(j) : d(i,j) \rightarrow [0,1]$.  The 
default neighbor-to-agent reward for $i$ is
\begin{align*}
    R^*_i = \frac{1}{|N(i)| - 1} \sum_{j \in N(i)/\{i\}} r^*_i(j).
\end{align*}
Policies can be
built around this metric.  For example, $i$ may choose to broadcast a hidden opinion (i.e., choose the action \emph{reveal} for all neighbors) as long as its average neighborhood opinion leans in the same direction as $i$'s private opinion.  We briefly describe a sample hand-crafted policy below.

Opinion updates are the penultimate event in a simulation, and by default, this mechanism acts as a simple majority rule: if a strict majority of $i$'s neighbors disagree with its opinion, then it will flip; otherwise, $i$'s opinion stays the same.  This is accomplished by setting $res(i) = 0,\ \Pr(upd_i) = 1\ \forall i \in V$, and $w_{ji} = 1\ \forall (j,i) \in E$.  For each node $i$, the system calculates the average opinion in $N(i)$ as in Eqn. \ref{aggregate}.  For instance, if $i$ has two friends with equal influence, then it will weight each of their opinions and its own at $1/3$ (described formally below).  This functionality can easily be augmented, e.g., to weight opinions differently based on their distance from one's own.\footnote{As seen in \cite{sirbu2013opinion}, where the distance between opinion vectors is related to their cosine similarity, or in \cite{chen2018modeling} which uses standard Euclidean distance.}  Keyword arguments allow for this value to be set uniformly for all agents (either set to 1 or another user-defined constant), or to be initialized randomly.  Once the aggregate opinion in a neighborhood is calculated, agents must decide what to do with it.  The most common method is to have agents become more similar to those in their neighborhood, so our software's default setting is to have agents tend toward agreement with their average neighborhood opinion.

We have implemented a custom archetype to emulate the overwhelmingly most common type of agent in social network modeling: the \emph{homogeneous} (HOM) archetype.  Such an agent prefers to be connected to others who hold the same opinions, and they get more reward from being connected to those who agree more.   Here we briefly describe the implementation.

As mentioned above, there are three aspects to an archetype: its reward function, its update mechanism, and its policy.  HOM agents prefer to be in like-minded neighborhoods, so their reward function should correlate positively with the amount of agreement they have with their neighbors.  The simplest way to accomplish this is to set $r_i^{hom}(j) = 1 - d(i, j)$.  Then, our agent will get more reward from more agreement, both on an individual basis and when considering the entire neighborhood.  This construction makes the HOM agent's goal clear: cause as much consensus as possible.  Secondly, we must implement an opinion update mechanism.  We use the most common construction for this archetype, in which
\[
b_{ik}^{t+1} =
\begin{cases}
-1(b_{ik}^{t}) & \mbox{ if } \widetilde{b}^t_{ik} * b_{ik}^{t} < 0 \\
b_{ik}^{t} & \mbox{ otherwise.}
\end{cases}
\]
The opinion flip happens probabilistically based on $\Pr(upd_i)$, setting which equal to 1 causes the flip to become deterministic.  Finally we describe the policy our HOM agent uses.  For our agent, this requires defining rules for unfriending and revealing an opinion.  The 
policy is: 

\noindent \textbf{foreach} neighbor $j \in N(i)$:

\textbf{if} $r^{hom}_i(j) < \mathit{unf\_thresh}$ \textbf{and} $j$ has revealed at least one

opinion to $i$,

\textbf{then} \emph{unfriend} $j$ with probability $\Pr(\mathit{unf})$

\textbf{else if} $\exists k : m_{ijk} = 0$, \emph{reveal} opinion $k$ with

probability 0.5. 

\noindent
This is a basic construction that allows our similarity-loving agent to excise any detrimental agents from its neighborhood, and take a chance on increasing agreement through revelation wherever there is an opportunity.

\section{Experiments} We conducted several experiments to validate our system, and to investigate the effects of different agent types on the evolution of a network's topology and opinion space.  We introduce two archetypes, \emph{heterogeneous} (HET) and \emph{adversarial} (ADV), in addition to the 
HOM type. HET agents seek opinion diversity in their neighborhoods, and 
get the most reward from an even split of agreements to disagreements with their neighbors; ADV agents actively seek out disagreement, and move away from their average neighborhood opinion rather than towards it.  We 
illustrate these archetypes' effect on the behavior of networks they inhabit.  In particular, we investigate: a) the effects of different proportional mixes of agent types, b) initial network densities, and c) resistance levels by agent type.

\begin{table}
\centering
\begin{tabular}{c|c}
 Name & Description \\ \hline
\emph{nodes} & number of nodes \\
$K$ & number of topics \\
\emph{type\_dist} & proportions for each archetype \\
\emph{saturation} & $\frac{1}{|V|} \sum_{i \in V} degree(i)$ \\
$\mathit{upd\_thresh}$ & strength of average opinion needed for a node to\\
& flip its opinion on a topic\\
$\mathit{upd\_prob}$ & probability that a node will flip its opinion on a\\
& topic if $update\_thresh$ is satisfied\\
$\mathit{unf}\_thresh$ & minimum pairwise reward needed to avoid unfriending\\
$\mathit{unf}\_prob$ & probability that a node will unfriend another if\\
& $update\_thresh$ is satisfied \\
$\mathit{friend\_prob}$ & probability of a potential edge being created \\ \hline
\end{tabular}
\caption{Simulation parameters}
\label{tab:paramlist}
\end{table}

Our ADV archetype is characterized by its desire for more disagreement to less, and their movement away from their neighborhood in opinion space rather than towards it.  This archetype has the reward function $r_i^{adv}(j) = d(i, j)$, and it obeys the update rule
\[
b_{ik}^{t+1} =
\begin{cases}
-1(b_{ik}^{t}) & \mbox{ if } \widetilde{b}^t_{ik} * b_{ik}^{t} > 0 \\
b_{ik}^{t} & \mbox{ otherwise.}
\end{cases}
\]
This update rule is similar to that used for partially antagonistic agents \cite{kurmyshev2011dynamics}, and the notion of nonconformity, or specifically wanting to have an opinion different from others, is reminiscent of the conformity parameter in \cite{duggins2014psychologically}.

The HET archetype follows the same update rule as HOM agents, but differs in the reward functions.  HET agents prefer to have a mix of agreement and disagreement with their neighbors, so their reward function is $r^{het}_i(j) = 1 - 2|d(i,j) - 0.5|$, peaking at a 50/50 agree-to-disagree ratio with a neighbor.  We are interested in this concept because it is clear that, while many people do in fact intentionally sequester themselves with like-minded groups, many actively seek out a diversity of opinions both contrary to and in line with their own in order to broaden their perspective.  We wish to investigate how the presence of such an agent type effects classical network outcomes.

Table \ref{tab:paramlist} lists all control parameters used in our simulations.  The following results were all obtained with the following constant simulation parameters: $\mathit{nodes} = 75$, $K = 4$, $\mathit{upd}\_\mathit{prob} = 0.25$, $\mathit{unf}\_\mathit{thresh} = 0.5$, $\mathit{unf}\_\mathit{prob} = 0.9$, $\mathit{friend}\_\mathit{prob} = 0.05$.  All networks were generated using the NetworkX small world algorithm ($\beta = saturation$).  We 
set all masks to visible 
to
investigate self-organization in networks with full observability.

\subsection{Model Verification}Here we briefly discuss our initial experiments to verify the baseline behavior of our model.

\noindent \textbf{Pure HOM Network:} Standard networks with homophilic agents experience a high degree of clustering and consensus.  Under the parameters above, pure HOM networks exhibited one of two behaviors in almost every instance: either the network would conglomerate into a single complete graph with one opinion consensus, or it would split early into two clusters (one usually much smaller than the other), each fully connected and with its own opinion consensus.

\noindent \textbf{Pure ADV Network:} Networks made only of ADV agents also tend to clump into just one or two groups early.  While HOM clusters eventually become completely connected, ADV clusters' density plateaus.  Also, ADV clusters either a) come to a unanimous consensus and then flip their opinion each time step, or b) form a core periphery structure with one opinion vector occupying the core and the opposite one filling the rest.

\noindent \textbf{Pure HET Networks:} HET agents do not like to deviate too far towards either complete agreement or disagreement with their neighbors.  When the agents obey a simple majority update rule, these networks always completely disconnect.  If HET agents are given sufficient resistance to opinion influence,\footnote{Empirically, ``sufficient" means that $\approx$75\% of one's neighbors must disagree before an opinion change is considered.} then the networks stay connected, but never reach maximum density, instead reaching a plateau like ADV networks.

\noindent \textbf{HOM/ADV Networks:} When divided evenly by type, these networks tend to remain connected throughout simulations, although complete self-sorting takes place.  They appear to split into two subgroups: the HOM one with complete consensus within, and the ADV one which again takes on a core periphery structure with two oscillating consensuses.

\noindent \textbf{HOM/HET Networks:} These networks separate consistently into a complete subgraph containing HOM agents, again at total opinion consensus, and HET agents eventually become isolates.  It appears that the prevalence of the HOM opinion draws HET towards it, but when consensus begins to form, HET agents remove their connections, even to each other.

\noindent \textbf{ADV/HET Networks:} These networks show the most ability to allow opinions to flow and also maintain topological fluidity rather than separation into static components.  They appear to reach a persistent equilibrium density after 20 to 50 steps.  These networks maintain the widest range of opinion representation.  The ADV cohort again forms a core periphery with two prevailing opinions.  The ultimate network density appears to depend directly on the resistance level of HET agents, but the diversity of opinion representation does not.

\subsection{Experiments on Fully Mixed Networks}The focal feature of our software is its attention to different agent types.  To the best of our knowledge, ours is the first investigation of truly different agent types interacting in a network together, rather than those using an attraction/repulsion model.  In particular we will lay out our initial observations about the HET archetype.  The following experiments were designed to elucidate the effects of proportional network composition, initial network density, and HET agent resistance levels at both the individual and network levels.
\begin{figure}
    \centering
    \includegraphics[width=0.9\columnwidth]{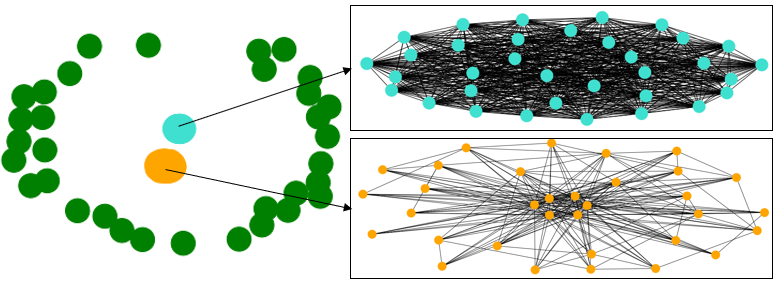}
    \caption{Fully mixed networks with uniform resistance to opinion influence separate into disjoint groups.  Types are HOM (blue), HET (green), and ADV (orange).  Node positions are arbitrary, but more densely connected clusters are shown physically closer than sparsely connected clusters.}\label{fig:separated}
\end{figure}
\subsubsection{The Effects of Network Composition}To investigate the impact of the agent type distribution on network outcomes, we ran 10 100-step simulations on networks with each of the following distributions (\% HOM / \% HET / \%ADV): 34/33/33, 50/25/25, 60/20/20, and 70/15/15.  For these experiments, we held $res(i) = 0$ for all agents, and $saturation = 0.15$.

Figure \ref{fig:separated} shows a typical outcome in these networks, independent of type distribution.  The network in the figure resulted from a 70/15/15 run, but the apparent patterns existed in all conditions tested.  Regardless of the distribution of agent types, these networks almost always separated into three cohorts: the HET agents, who end up isolated in the network as consensus begins to take over in the core; the HOM agents, who again aggregate into a complete subgraph (unless their numbers were great enough, in which case they typically split into two disjoint, disagreeing clusters); and the ADV agents form a core periphery cluster.   Further, in each simulation the network would remain without isolates for several steps until one HET agent left; once that happened, the rest of the HET agents left very quickly thereafter.  The ADV and HOM clusters, though already formed, never separate from each other until most or all of the HET agents leave.

These experiments showed that each of these agent types continues to induce characteristic network outcomes as laid out above (e.g. HOM agents cluster and arrive at unanimous opinions or split into two separate camps, ADV agents form their own core periphery, etc.), but those impacts have some interplay.  The importance of HET agents in keeping the other groups connected in these experiments showed that some types of agents can mediate the interplay between others.
\begin{figure}
    \centering
    \includegraphics[width=0.95\columnwidth]{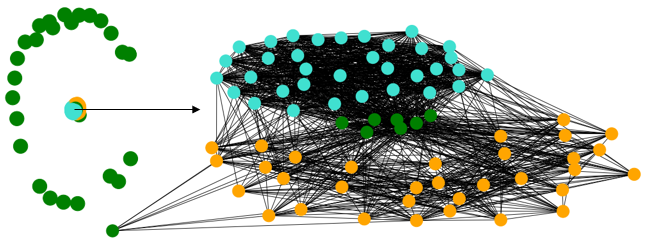}
    \caption{When HET agents have a higher resistance to influence, they prevent networks from separating into disjoint camps.}
    \label{fig:glued}
\end{figure}
\subsubsection{The Effects of Initial Density}The initial density of the network can have significant influence over its evolution.  For example, in pure ADV networks an initial density set too small will cause the network to fragment more.  We used the same conditions as in the previous experiments, but with $saturation = 0.05, 0.1, 0.15, 0.2,$ and $0.25$.

At a saturation of 0.05, agents did not have enough connections to form anything more than two- to six-member components under any type distribution.  However, more evenly-split networks did tend to break apart into smaller groups on average.  ADV agents also appeared to benefit from this slightly, in some cases maintaining a cluster with one of the other agent types.

An overly sparse initial network lends itself to fragmentation as would be expected, whereas sufficiently dense starting networks appear to almost always separate into disjoint components delineated by agent type.  The novel illustration is that \emph{polarization is not restricted to opinion space}.  In other words, in our experiments we saw not only the expected polarization in opinion --- as with the core periphery tendency of ADV agents or the multi-clustering behavior of HOM agents --- but also in the proclivity of agents to self-segregate based on type.  Even HET agents, before they split apart entirely, formed tighter communities with each other than with either of the other archetypes.  This means that even non-homophilic agents tend to bond most strongly and persistently with each other.
\begin{figure}
    \centering
    \includegraphics[width=0.8\columnwidth]{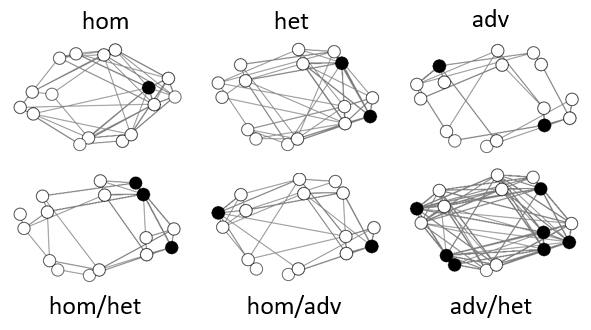}
    \caption{Movements through opinion space with different archetypes.  White dots represent initial opinions, black dots are terminal opinions, and lines are agent paths over time.  Most single- or dual-type networks tend towards a small set of terminal opinions, but ADV/HET networks often maintain more opinion diversity.}\label{fig:ops}
\end{figure}
\subsubsection{The Effects of HET Agent Resistance Levels}Our main point of inquiry in this work is: how does the presence of HET agents affect the network?  The observations of our last two sets of experiments make clear that agents seeking balance may have additional complications finding a suitable situation for themselves within the network given the behavior of other archetypes.  These agents also seem to have a cohesive effect on the network as a whole.  Whenever mixed networks split apart along opinion and/or archetype lines, they only do so after most of the HET agents have left.

It makes sense that these agents would choose to leave the network once the other types have entrenched themselves in their own segregated camps, because by definition HET agents have two reward ``valleys": total agreement and total disagreement.  These two valleys overlap the reward peaks of the other two archetypes, creating a balancing act between them.  Our last set of experiments is designed to test the aggregate effects of endowing HET agents with greater resistance to opinion influence.
\begin{figure}
    \centering
    \includegraphics[width=0.85\columnwidth]{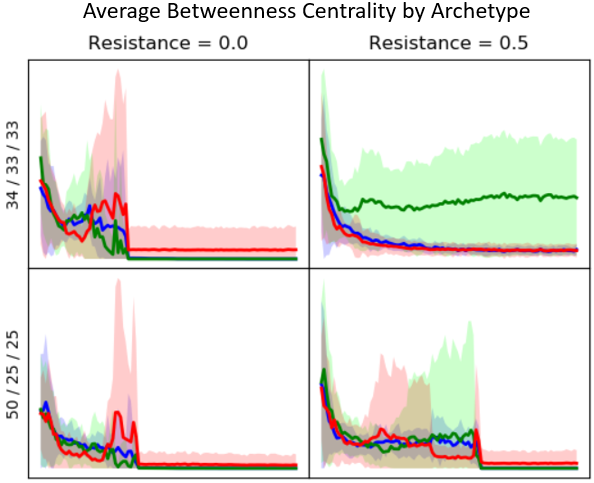}
    \caption{The effects of resistance to opinion change on betweenness centrality.  \emph{Top row:} All three archetypes evenly represented; HOM (blue), HET (green), ADV (red). \emph{Bottom row:} HOM agents twice as common as others.  
    \emph{Columns:}  HET agents susceptible/resistant to opinion influence.}\label{fig:diffs}
\end{figure}
Agents in this style environment who have a greater openness to different opinions tend to foster consensus rather than hinder it \cite{motsch2014heterophilious}.  We varied the value of $res(i)$ for all HET agents from 0.0 to 0.5, which corresponds to agents needing between half and 3/4 of their neighbors to disagree with them on a topic before they might flip their opinion.

In most tests with $res(i)$ set to 0.0 for HET agents, the networks split apart into camps.  Figure \ref{fig:glued} illustrates the characteristic outcome we observed when we increased that value to 0.25.  After 100 steps, most of the HET agents in the network had left, just as before.  However, a small cluster of them remained in between the two other clusters, which organized themselves in their characteristic ways.  It can be seen that one agent was possibly about to leave the network, even after 100 steps, so it is immediately evident that HET agents alleviate some topological rigidity.

Networks composed of only homophilic agents tend toward at most two different opinion poles, even for multidimensional opinion spaces.  The introduction of ADV agents increases this number of terminal opinion states by an additional two, but only exceptional circumstances violate this rule of thumb.  Figure \ref{fig:ops} shows examples of agents' movements through opinion space under different type compositions.  The white dots represent each agent's starting point in opinion space, and the black dots represent each agent's terminal opinion.  The lines connecting them pass through all opinions adopted in between.

Since HET agents appear to have significant structural effects on the network, we also explored how agents performed under our test conditions according to several different criteria.  We collected several centrality statistics and stepwise reward for each agent throughout our experiments.  Figure \ref{fig:diffs} illustrates the effect of HET agent resistance on networks of two different compositions.  We chose to present betweenness centrality scores because of the connecting role HET agents appear to play.

Each column in the figure corresponds to a single setting of HET resistance, and each row corresponds to a network composition --- the top row is evenly divided among archetypes, and
the bottom row has twice as many HOM agents as HET or ADV agents.  The left column shows a typical outcome in networks with HET agents that have just as low a threshold for opinion change as everyone else.  Lines in the figure represent the average score across all agents of a given type, and shaded areas represent the mean $\pm$ one standard deviation.

The left third of each plot shows agents' betweenness centrality during the initial phase of the simulation when agents are testing out the most connections and trimming their neighborhoods to suit them.  There is a peak in ADV betweenness, generally around step 40, and one for HET agents as well around the same time, before a steep drop off.  This sudden decline corresponds to the moments leading up to and including the separation of subclusters from each other.  The right column shows the same metric when HET agents have significant resistance to opinion influence.  In these two plots it can be seen that HET agents' betweenness centrality is greatly enhanced when outside influence means less to them.  In the evenly split network, most HET agents were never separated from the rest of the network, although this agent type had the most variation across individuals.  As in Figure \ref{fig:glued}, these agents kept almost the entire network in one piece.  In the 50/25/25 network, HET agents still eventually separated from the network most times, but they were able to keep it together for a longer span than when they were more fickle with their opinions.  Further, the plots demonstrate that resistant HET agents were the last connectors of the network's subcomponents, empirically validating our observations from the second round of experiments.



\section{Conclusion}
In this paper, we presented a software suite in ongoing development designed to support a broad range of diffusion simulations, and especially opinion diffusion in an environment with agents that can choose which opinions to reveal to whom, and whom to unfriend.  We used this platform to investigate novel agent types not yet present in the literature, and explored their effects on network outcomes, opening the door to new research extensions.

\bibliographystyle{plain}
\bibliography{main.bib}

\end{document}